# Comparative Fault Location Estimation by Using Image Processing in Mixed Transmission Lines

[1]Serkan BUDAK, [2]Bahadır AKBAL

*[1]Konya Teknik Üniversitesi, Mühendislik ve Doğa Bilimleri Fakültesi, Elektrik-Elektronik Mühendisliği Bölümü, KONYA*
*[2]Konya Teknik Üniversitesi, Mühendislik ve Doğa Bilimleri Fakültesi, Elektrik-Elektronik Mühendisliği Bölümü, KONYA*
[1] sbudak@ktun.edu.tr, [2] bakbal@ktun.edu.tr

**ABSTRACT:** Overhead lines are generally used for electrical energy transmission. Also, XLPE underground cable lines are generally used in the city center and the crowded areas to provide electrical safety, so high voltage underground cable lines are used together with overhead line in the transmission lines, and these lines are called as the mixed lines. The distance protection relays are used to determine the impedance based fault location according to the current and voltage magnitudes in the transmission lines. However, the fault location cannot be correctly detected in mixed transmission lines due to different characteristic impedance per unit length because the characteristic impedance of high voltage cable line is significantly different from overhead line. Thus, determinations of the fault section and location with the distance protection relays are difficult in the mixed transmission lines. In this study, 154 kV overhead transmission line and underground cable line are examined as the mixed transmission line for the distance protection relays. Phase to ground faults are created in the mixed transmission line. overhead line section and underground cable section are simulated by using PSCAD/ EMTDC ™. The short circuit fault images are generated in the distance protection relay for the overhead transmission line and underground cable transmission line faults. The images include the R-X impedance diagram of the fault, and the R-X impedance diagram have been detected by applying image processing steps. Artificial neural network (ANN) and the regression methods are used for prediction of the fault location, and the results of image processing are used as the input parameters for the training process of ANN and the regression methods. The results of ANN and regression methods are compared to select the most suitable method at the end of this study for forecasting of the fault location in transmission lines.

## Karma İletim Hatlarında Görüntü İşleme Kullanılarak Karşılaştırmalı Hata Konumu Tahmini

**ÖZ:** Havai hatlar genellikle elektrik enerjisi iletimi için kullanılır. Ayrıca XLPE yeraltı kablo hatları genellikle şehir merkezinde ve kalabalık alanlarda elektrik güvenliğini sağlamak için kullanılır, bu nedenle iletim hatlarında havai hat ile birlikte yüksek gerilim yeraltı kablo hatları kullanılır ve bu hatlar karma hatları olarak adlandırılır. Mesafe koruma röleleri, iletim hatlarındaki akım ve gerilim büyüklüklerine göre empedans tabanlı ölçüm sonucu arıza yerini belirler. Ancak yüksek gerilim kablo hattının karakteristik empedansı havai hattan önemli ölçüde farklı olduğundan, birim uzunluk başına farklı karakteristik empedans nedeniyle karma iletim hatlarında arıza konumu doğru bir şekilde tespit edilemez. Bu nedenle karma iletim hatlarında mesafe koruma röleleri ile arıza bölümünün ve yerinin tespiti zordur. Bu çalışmada, 154 kV havai iletim hattı ve yer altı kablo hattı, mesafe koruma röleleri için karma iletim hattı olarak incelenmiştir. Karma iletim hattında faz-toprak arızaları oluşturulur ve havai hat bölümü ve yeraltı kablo bölümü PSCAD / EMTDC ™ kullanılarak benzetimi yapılmıştır. Kısa devre arıza görüntüleri, havai iletim hattı ve yer altı kablo iletim hattı arızaları için mesafe koruma rölesinde oluşturulur. Görüntüler, arızanın R-X empedans diyagramını içerir ve R-X empedans diyagramından elde edilen görüntüler görüntü işleme adımları uygulanmıştır. Arıza yeri tahmini için görüntü işleme sonuçlarından çıkarılan özellikler giriş parametresi olarak belirlenmiştir. Yapay sinir ağları (YSA) ve regresyon yöntemleri kullanılarak arıza yeri tahmini yapılmıştır. YSA sonuçları ve regresyon yöntemleri bu çalışmanın sonunda iletim hatlarında arıza yerinin tahmin edilmesi için en uygun yöntemin seçilmesi için karşılaştırılmıştır.

*Anahtar Kelimeler:* *Mesafe koruma rölesi, Karma iletim hatları, Kısa devre arızaları, Regresyon metotları, Arıza yeri tahmini, Yapay sinir ağları*

## INTRODUCTION

Protection systems are one of the most important parts of power systems. Inside the protection and control systems used in the transmission of electrical energy, the most important task falls on protection relays.

Various places and various short circuit faults can occur in electrical power systems. When the causes of short circuit faults are examined, lightning strikes, high-speed winds, ice weight, earthquakes, fires, explosions, falling trees, flying objects, animals, people and other natural cases, as well as hardware failure, hardware aging, wrong hardware, wrong operation, wrong switching such as can cause. Protective relays are installed at different points and parts of the power system to make fast and selective protection.

Distance protection relays are one of the most preferred relays to protect transmission lines (Glover, Sarma, & Overbye, 2012). While distance protection relays are applied as main protection in transmission and distribution lines, they are widely used as backup protection for transformer, busbars and distant lines. Distance protection relays are adjusted to operate according to the impedance of the line by comparing the voltage and current values of the transmission line (Glover et al., 2012). The distance of the point where the relay is connected to the fault is determined by measuring the impedance value. Short circuit impedance decreases at a certain rate as it approaches the feeding point of the line and thus the fault location is determined. Determining the fault location in transmission lines is an important feature used in protection systems. The increasing complexity of power transmission systems has greatly increased the importance of fault location investigation studies in recent years. If the location of a fault is known or can be estimate with high accuracy, the fault can be repair quickly. Quick troubleshooting is of great importance as it reduces customer complaints, downtime, operating cost, loss of revenue, and maintains the stability of the system.

Overhead line systems are dominant in the transmission of electrical energy. Overhead line systems are transmission lines that have proven their operational reliability and functional use. In addition, the installation costs of overhead lines are low and their useful life is high (Han & Crossley, 2013).

Nowadays, underground cables are used in distribution lines, especially in places where there is a high density of people, in order to ensure operational safety, human-environmental health and the economy of the enterprise. With new technological developments in insulation systems in high voltage XLPE underground cables, cables have been used in transmission lines. Underground power cable installations have started to replace some of the overhead transmission lines due to environmental factors in areas where the overhead line cannot pass and in densely populated areas (Han & Crossley, 2013).

When the operations of the distance protection relay in mixed transmission lines (lines consisting of overhead and underground cable sections) were examined, it was seen that it could not detect the correct fault location. Pinpointing the location of fault on transmission lines can significantly save labor force and speed up the repair process. For this reason, efforts are being made to use mixed transmission lines with high efficiency (Han & Crossley, 2013).

When the studies in the literature are examined, machine learning-based approaches (Ekici, 2012), (Livani & Evrenosoglu, 2013), (Thukaram, Khincha, & Vijaynarasimha, 2005), (Khorashadi-Zadeh, 2004) and (Fan, Yin, Huang, Lian, & Wang, 2019) are also suggested to find transmission line failures. In Ekici, proposes a support vector machine (SVM) to predict the transmission line fault location. It used a wavelet transform to process single-ended fault transient data and then used it as an input data set to the SVM network. In Livani et al, used the SVM network to predict fault locations with a separate wavelet transform to extract features from waves moving on arrival. In Thukaram et al, an ANN and SVM approach is presented to find errors in radial distribution systems. Unlike traditional fault section estimation methods, the proposed approach uses the measurements available at the substation, circuit breaker and relay situations. In Khorashadi et al, A new scheme has been proposed for fast and reliable fault classification for a double-circuit transmission line. It has been shown that the algorithm can work quickly and accurately in different fault conditions such as fault type, fault resistance, fault initiation angle, fault location, pre-fault power flow direction and system short circuit level. In Fan et al, they offer a new single-ended fault location approach for transmission lines using modern deep learning techniques. They estimate the given error distance using single-ended voltage and current measurements.

Different techniques have been proposed in (Aziz, khalil Ibrahim, & Gilany, 2006), (Niazy & Sadeh, 2013) and (Han & Crossley, 2013) when determining the fault location in mixed transmission lines. In Aziz et al, for transmission systems consisting of an overhead line coupled with an underground cable, a fault location diagram is proposed. Requires phasor synchronous measurements from both ends of the transmission line. In Niazy et al, A new single-ended fault detection method has been proposed for underground cable coupled with overhead lines. In the proposed method, samples are taken from the transient voltages generated by the fault clearing action of the circuit breaker from the sending end of the cable line. When the wavelet transform is applied, the waves going to the fault detector are detected. The fault section, the overhead line section or the underground cable section is identified, and then the wave velocity is calculated and the fault location is accurately determined. In Han et al, proposed a new hybrid fault location scheme for fault section detection in a mixed transmission line (MTL) consisting of an overhead line (OHL) and an underground cable (UGC).

In this study, unlike the studies performed on mixed transmission lines, the fault location was estimated by using the R-X impedance diagram images of the distance protection relay at the time of short circuit by using ANN and Regression methods. ANN and Regression methods were compared with each other and the best method was determined. The use of image processing techniques and their comparison with different methods differs from other studies.

**MATERIALS AND METHOD**

Distance protection relays are used to protect transmission lines against short circuit faults and to determine the location of the fault. Since distance protection relays make impedance-based fault and fault location detection according to current and voltage magnitudes, it detects fault location in mixed transmission lines due to different impedance per unit length (Han & Crossley, 2013, 2015). In this study, images taken from the R-X impedance diagram were applied to image processing steps for short circuit faults occurring in the overhead transmission line and underground cable transmission line. Data sets were created to be used in ANN and Matlab Regression Learner applications. By comparing the performance results of ANN and Regression methods, fault location estimates are given in the results section of the study.

**Mixed Transmission Lines**

The underground cable line can replace a part of the existing overhead line in high voltage transmission lines, but the characteristic impedance of the cables differs according to the overhead line. The inductance of underground cable conductors is 30-50% less than an overhead line and its capacity is 30-40 times greater than an overhead line (Tziouvaras, 2006). Using the underground cable line and the overhead line together makes it difficult to determine the fault section and location in the transmission line (Tziouvaras, 2006).

Identifying the fault location accurately and quickly will reduce the costs of locating the fault. It will speed up the operations that need to be done quickly, such as repairing the fault and re-commissioning the transmission line, and reduce production, usage and income losses due to interruptions. For this reason, accurate and fast estimation and determination of the malfunction is very important in terms of operating continuity of the system, operating cost and operational safety.

**Image Processing Method**

Image processing is a method applied to obtain different information using different techniques over digitally transferred images. Image is used mathematically in the design and analysis of image processing systems. In the image processing method, many different information can be obtained that we cannot see with the eye.

When images are digitized using image processing method, a large size matrix is obtained. Using the large size matrix has a high error rate and long processing time during the training stage. For this reason, by applying various statistical methods, the matrix size is reduced and the processing time and error rate are reduced. This process is done by extracting the features of the images before proceeding to the training phase.

In this study, some statistical approaches used in feature extraction are used. With the Gray Level Co-occurrence Matrices (GLCM) function, we can obtain different statistical properties for each image. These are mean, entropy, variance, difference, contrast, inverse difference moment, energy, correlation, cluster shade, cluster prominence, sum entropy, sum mean, difference entropy, sum variance values. The training phase was started by extracting the statistical properties of 1x20 size images for each image (Budak & AKBAL).

The training data set was created by using images image processing methods. The applied image processing steps are shown in Figure 1 (Budak & AKBAL).

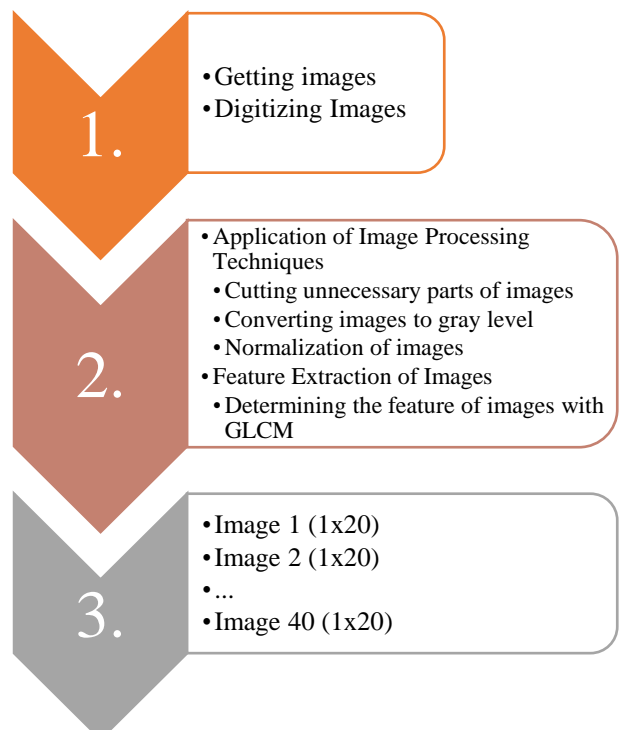

**Figure 1.** Applied image processing steps

**Artificial Neural Networks**

ANN emerged as a result of similarity with the working system of the human brain. It is an algorithm that resembles the working system of the human brain, which has abilities such as learning, associating, decision making, inference, recall, generalization, classification, new information generation, feature determination, optimization in computer environment.

In Figure 2, the general structure of ANN consists of three layers. These are the input layer, hidden layer, and output layer. The input layer has neurons that receive input from outside. Neurons in this layer transmit the input values to the hidden layer. A neuron in the hidden layer receives signals from all the neurons of the previous layer. After processing these neurons, it sends its output to all the neurons of the next layer. Output layer is a layer that contains neurons that transmit the outputs out. Neurons in the output layer process information from the hidden layer and output (Yegnanarayana, 2009).

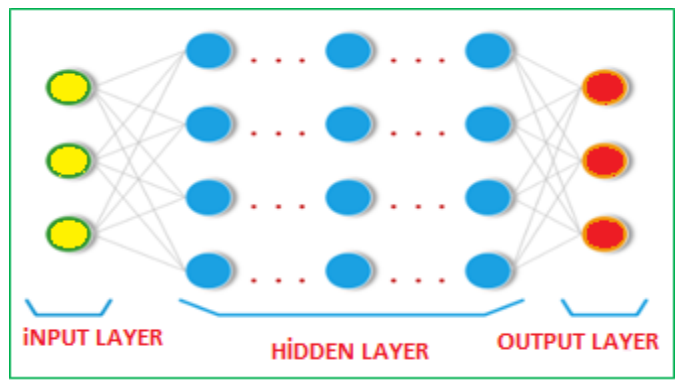

**Figure 2.** General structure of ANN

In ANN, the way the problem is given to the network, the structure of the network, the algorithm used, the learning strategy used by the network, the transfer function and learning rules affect the performance of the network (Öztemel, 2012). In this study, various network types, learning functions, training functions and performance functions in ANN were tried to determine the most suitable network structure for the problem.

**Matlab Regression Learner App**

Regression Learner is an application that provides training, comparison and determination of the most appropriate method using machine learning methods widely used in the literature (MathWorks). Input and output data sets are given to the application, data is checked, features are selected, verification schemes are determined and performance results are obtained by training with different models. The results for each training model can be visualized as a Response Plot, Predicted vs. Actual Plot and Residual Plot. Performance criterion looked at to examine and evaluate the predictive accuracy of training models. Performance criterion determined the performance of the model and allow us to choose the best suitable model. The application also provides information on estimated training time and speed.

The training models used in Matlab Regression Learner application are Linear Regression, Tree, Ensemble Trees, Support Vektor Machines (SVM), Gaussian Process Regression and their sub-methods (MathWorks). The best training model can be found automatically. In this way, there is no obligation to write separate and different code. It has the ability to automatically train and test multiple models at the same time (Özleyen, 2019).

**Simulation Study**

In the PSCAD ™ / EMTDC ™ simulation program, a mixed transmission line was formed by modeling 154 kV, 50 Hz, 200 km and 50 km two overhead transmission lines and 10 km of underground cable lines between the overhead lines. The transmission line is fed from two generator generation sources connected to two three-phase power transformers. For the overhead transmission line, 154 kV single circuit power transmission line, 1272 MCM conductor cross sections and 'PB' pole type are designed. 89/154 kV, 2XS(FL)2Y cable type is designed for underground cable transmission line. The low voltage side of the transformer has been selected as 11 kV delta, and the high voltage side as 154 kV star. A fixed load of 20 MW was used in the system (Budak, 2020).

In the study, the images of short circuit faults occurring both in the overhead transmission line and in the underground cable transmission line were observed in the R-X impedance diagram. Short circuit fault impedance is determined as constant $Z_f=1\Omega$. Short circuit fault persists for 0.05 seconds and 0.3 seconds after the system operates for fixed periods. Studies have been carried out by creating a phase a-ground short circuit, which is the most common single-phase ground fault. Figure 3 shows the mixed transmission line model.

Figure 2 shows sample images occurring in phase a-ground short circuit faults formed at the 60th km of the overhead line section of the mixed transmission line and at the 5th km of the underground cable line section.

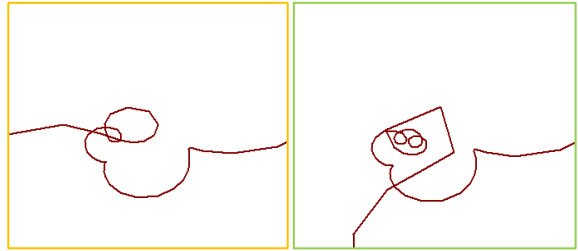

**Figure 3.** Sample images obtained from the R-X impedance diagram in mixed transmission lines

Root Mean Square Error (RMSE) and percentage error value were used to evaluate the results obtained in fault location estimation studies. The RMSE and error value are always positive, and close to zero indicates the best value. Equation 1 and Equation 2 contain percent error and RMSE equations, respectively (Karasu, Altan, Saraç, & Hacıoğlu, 2018).

$$\%Error\ Value = \left| \frac{Actual\ fault\ location - Calculated\ fault\ location}{Total\ length\ of\ the\ line} * \right|100 \quad (1)$$

$$RMSE = \sqrt{\frac{\sum_{i=1}^{n}(xi-yi)^2}{n}} \quad (2)$$

**RESULTS**

40 number phase a-ground faults were created at 5,10,15,… 200 km of the overhead transmission line simulated using PSCAD ™ / EMTDC ™ and 50 number phase a-ground faults were created at 0.2, 0.4,… 10 km of the underground cable transmission line.

In order to determine the fault location, three different network types, namely Feed Forward Backpropagation, Cascade Feed Forward Backpropagation and Elman Feedback Network, and five different training functions LM, CGB, OSS, GDX and SCG were used in the ANN. In the ANN model, the number of hidden layers, the number of neurons in hidden layers and which activation function will be used in hidden layers were determined by trial and error method. The tangent sigmoid activation function is used in the input and hidden layers of the specified network, and a linear activation function in the output layer. The number of neurons in the layers was determined as 1, 10 and 1, respectively. The iteration number was chosen as 1000. Of the 40 faults in the overhead transmission line data set, 32 were used in the training set, 8 in the test set and of the 50 faults in the underground cable transmission line 40 were used in the training set and 10 were used in the test set.

In Table 1, the short circuit faults occurring in the overhead line and underground cable line are simulated and the methods and training errors used in the ANN are given (Budak, 2020).

**Table 1.** Methods and training errors used in ANN application

| Method | | RMSE OHL | RMSE UGC |
|---|---|---|---|
| Feed Forward Backpropagation | LM | **0.0048** | 0.0082 |
| | CGB | 0.0106 | 0.0123 |
| | OSS | 0.0269 | 0.0176 |
| | GDX | 0.0124 | 0.0113 |
| | SCG | 0.0153 | 0.0098 |
| Cascade Feed Forward Backpropagation | LM | 0.0096 | 0.0139 |
| | CGB | 0.0106 | 0.0162 |
| | OSS | 0.0411 | 0.0159 |
| | GDX | 0.0400 | 0.0144 |
| | SCG | 0.0163 | 0.0151 |
| Elman Feedback | LM | 0.0062 | 0.0129 |
| | CGB | 0.0121 | 0.0135 |
| | OSS | 0.0146 | 0.0126 |
| | GDX | 0.0157 | 0.0110 |
| | SCG | 0.0143 | 0.0120 |

In Table 2, Fault location prediction training errors in overhead and underground cable transmission lines are given by using regression methods.

**Table 2.** Regression methods used and training errors

| Method | | RMSE OHL | RMSE UGC |
|---|---|---|---|
| Linear Regression | Linear | 0.017721 | 0.010085 |
| | Interactions Linear | 0.025492 | 0.0064932 |
| | Robust Linear | 0.036589 | 0.010119 |
| | Stepwise Linear | 0.026962 | **0.0060709** |
| Tree | Fine Tree | 0.098382 | 0.061107 |
| | Medium Tree | 0.15016 | 0.12531 |
| | Coarse Tree | 0.23998 | 0.24734 |
| SVM | Linear SVM | 0.02837 | 0.016424 |
| | Quadratic SVM | 0.056601 | 0.020054 |
| | Cubic SVM | 0.21514 | 0.03174 |
| | Fine Gaussian SVM | 0.12044 | 0.071415 |
| | Medium Gaussian SVM | 0.065541 | 0.022517 |
| | Coarse Gaussian SVM | 0.06412 | 0.034763 |
| Ensemble Trees | Boosted Trees | 0.051223 | 0.046086 |
| | Bagged Trees | 0.075814 | 0.058129 |
| Gaussian Process Regression | Spuared Exponential GPR | 0.0078464 | 0.0092204 |
| | Matern 5/2 GPR | 0.010683 | 0.0087294 |
| | Exponential GPR | 0.023817 | 0.0093471 |
| | Rational Quadratic GPR | 0.0079423 | 0.0094911 |

Looking at the training errors obtained from ANN and Regression methods, the methods that give the best results were chosen. In Table 3 and Table 4, The estimation of fault locations are given as a result of short circuit fault occurring at the determined kilometers of the mixed transmission line and the percentage error values of these estimates.

**Table 3.** Estimated (km) fault locations and percentage error value as a result of short circuit occurring in overhead transmission line

| Method | Fault Location (km) | Estimated Fault Location (km) | Error Value (%) |
|---|---|---|---|
| Feed Forward Backpropagation | 20 | 19.1112 | 0.444 |
| | 45 | 42.4392 | 1.280 |
| | 70 | 70.3472 | 0.1736 |
| | 95 | 94.8647 | 0.0676 |
| | 120 | 119.8546 | 0.0727 |
| | 145 | 143.1658 | 0.9171 |
| | 170 | 169.8743 | 0.0628 |
| | 195 | 195.0882 | 0.0441 |

**Table 4.** Estimated (km) fault locations and percentage error value as a result of short circuit occurring in underground cable transmission line

| Method used | Fault Location (km) | Estimated Fault Location (km) | Error Value (%) |
|---|---|---|---|
| Linear Regression Stepwise Linear | 0.8 | 0.835775 | 0.357 |
| | 1.8 | 1.707975 | 0.921 |
| | 2.8 | 2.687975 | 1.121 |
| | 3.8 | 3.7623 | 0.377 |
| | 4.8 | 4.817025 | 0.17 |
| | 5.8 | 5.7321 | 0.679 |
| | 6.8 | 6.720675 | 0.794 |
| | 7.8 | 7.757025 | 0.43 |
| | 8.8 | 8.66965 | 1.304 |
| | 9.8 | 9.755 | 0.45 |

Figure 5 and Figure 6 show the graphics of the actual and estimate fault locations of the overhead and underground cable lines, respectively.

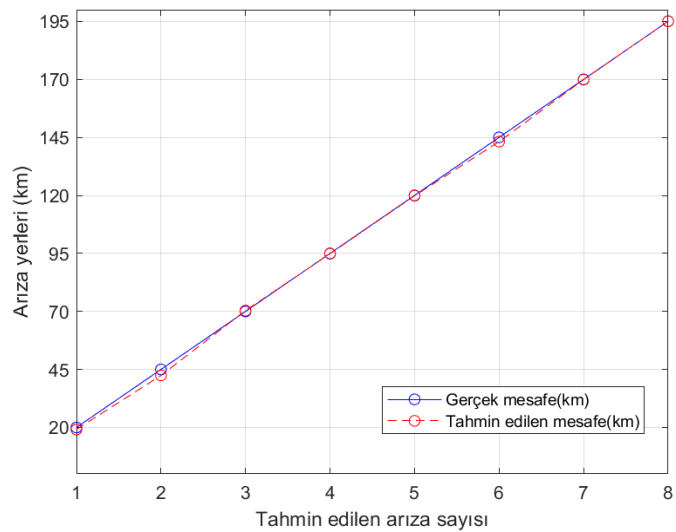

**Figure 5.** Comparison graph of actual and estimated distances in overhead transmission lines

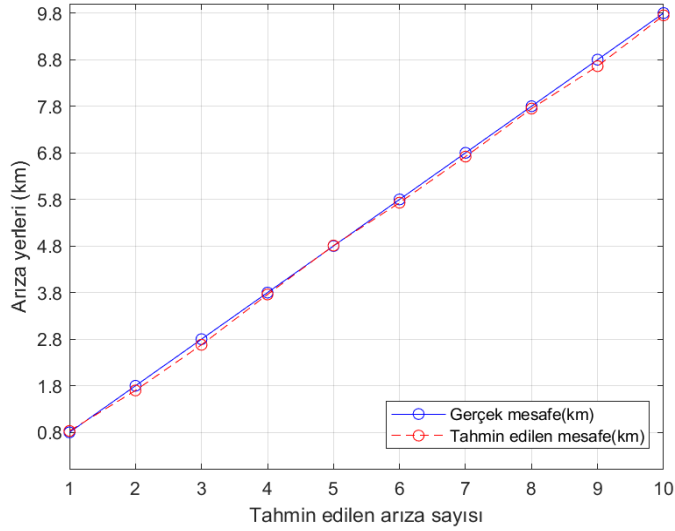

**Figure 6.** Comparison graph of actual and estimated distances in underground cable transmission lines

## CONCLUSION

In this study, image processing is used for short circuit faults occurring in mixed transmission lines. The fault location was estimated using ANN and Regression methods. Considering the methods and training errors used as a result of the studies carried out on the mixed transmission line, it has been determined that the best training model is the Feed Forward Back Propagation LM network type in the overhead transmission line section. When we look at the underground cable line section, it was determined that the best training model was Linear Regression (Stepwise Linear). According to these results, it has been seen that better results can be obtained by using different methods.

According to the test results, the highest fault location estimation is 1.28% in the estimated distances in the overhead line section and 1.304% in the underground cable line part. In the study, fault location predictions can determine whether the fault is in the overhead line or underground cable line part in the case of short circuit failures occurring in the mixed transmission line and close to the actual fault location. Unlike other studies, high predictive values were obtained in different methods using image processing. The use of image processing in protection systems shows that it can be improved and used in different ways.

As a result, in long mixed transmission lines, in case of short circuit failure, the mixed transmission line does not affect fault location estimation studies and high predictive values are shown in the tables.